\documentclass[aps,prd,twocolumn,nofootinbib,superscriptaddress]{revtex4-1}
\usepackage[utf8]{inputenc}
\usepackage{graphicx}
\usepackage{subfigure}
\usepackage{datatool}
\usepackage{amsmath,amssymb}
\usepackage[scientific-notation=true]{siunitx}
\usepackage{lineno}
\usepackage{natbib}
\usepackage[dvipsnames]{xcolor}
\usepackage{xspace}
\usepackage{orcidlink}

\DTLsetseparator{,}

\def\gwh{gravitational-wave\xspace}
\def\gws{gravitational waves\xspace}
\def\dm{dark matter\xspace}
\def\dmh{dark-matter\xspace}

\def\uldm{ultralight dark matter\xspace}
\def\uldmh{ultralight dark-matter\xspace}

\newcommand{\TFFT}{T_\text{FFT}}

\newcommand{\Tobs}{T_\text{obs}}

\newcommand{\bea}{\begin{eqnarray}}
\newcommand{\eea}{\end{eqnarray}}
\newcommand{\be}{\begin{equation}}
\newcommand{\ee}{\end{equation}}

\newcommand{\rhoDM}{\rho_{\text{DM}}}

\newcommand{\tcoh}{T_\text{coh}}

\newcommand{\lpf}{LISA Pathfinder\xspace}

\newtoggle{fullauthorlist}
\toggletrue{fullauthorlist}
\newtoggle{endauthorlist}
\toggletrue{endauthorlist}

\begin{document}

\title{First search for ultralight dark matter with a space-based gravitational-wave antenna: LISA Pathfinder }

\author{Andrew L. Miller\,\orcidlink{0000-0002-4890-7627}}
\email{andrew.miller@nikhef.nl}
\affiliation{Université catholique de Louvain, B-1348 Louvain-la-Neuve, Belgium}
\affiliation{Nikhef -- National Institute for Subatomic Physics,
Science Park 105, 1098 XG Amsterdam, The Netherlands}
\affiliation{Institute for Gravitational and Subatomic Physics (GRASP),
Utrecht University, Princetonplein 1, 3584 CC Utrecht, The Netherlands}

\author{Luis Mendes}
\email{Luis.Mendes@ext.esa.int}
\affiliation{RHEA Group for ESA, Camino bajo del Castillo s/n, Urb. Villafranca del Castillo, Villanueva de la Cañada, 28692 Madrid, Spain}


\begin{abstract}
We present here results from the first-ever search for dark photon dark matter that could have coupled to baryons in LISA Pathfinder, the technology demonstrator for a space-based gravitational-wave antenna. After analyzing approximately three months of data taken by LISA Pathfinder in the frequency range $[2\times 10^{-5},5]$ Hz, corresponding to dark photon masses of $[8\times 10^{-20},2\times 10^{-14}]$ eV/$c^2$, we find no evidence of a dark-matter signal, and set upper limits on the strength of the dark photon/baryon coupling. To perform this search, we leveraged methods that search for quasi-monochromatic gravitational-wave signals in ground-based interferometers, and are robust against non-Gaussianities and gaps in the data. Our work therefore represents a proof-of-concept test of search methods in LISA to find persistent, quasi-monochromatic signals, and shows our ability to handle non-Guassian artifacts and gaps while maintaining good sensitivity compared to the optimal matched filter. The results also indicate that these methods can be powerful tools in LISA to not only find dark matter, but also look for other persistent signals from e.g. intermediate-mass black hole inspirals and galactic white dwarf binaries.
\end{abstract}

\maketitle
\section{Introduction}\label{sec:intro}

\lpf \cite{McNamara:2008zz,Armano:2009zz,Antonucci:2012zz} was a demonstrator of some of the technologies to be used in LISA, the space-based \gwh antenna to be launched in the second half of the next decade \cite{danzmann2003lisa,LISA:2017pwj}. Its main goal, the demonstration that the noise in the separation of two test masses $\approx$ 40cm apart could be kept at a suitably low level to allow the detection of gravitational waves by LISA, was successfully achieved by \lpf. In fact, \lpf was a remarkable success and produced a noise power spectral density that surpassed the LPF mission requirement by more than an order of magnitude for some frequency regions and by a factor of a few compared to the LISA requirement \cite{Armano:2016bkm,Armano:2018kix}.


Though designed for the purpose described above, the data from \lpf have been used in different ways to probe other areas of physics, such as quantifying possible noise correlations in future stochastic \gwh background searches \cite{Boileau:2022uos}, measuring the value of Newton's gravitational constant \cite{LISAPathfinder:2019bgi}, testing the strong equivalence principle when spacecrafts orbit Lagrange points \cite{Congedo:2016mlx}, and bounding collapse models \cite{Carlesso:2016khv}. Such studies show that high-precision measurements of acceleration or displacement can be employed to tackle interesting physics problems for which the experiment was not primarily designed, and motivate the need for further analyses of \lpf data \cite{lpfarchive} to see what other kinds of questions that this data, and the future LISA mission, can answer.

\lpf data can be used to probe the existence of \uldm directly via its interactions with the test masses. Essentially, the instruments exists in a field or ``wind'' of \dm, and this \dm would couple to standard model particles in the test masses, causing sinusoidal oscillations of their positions over time at a frequency fixed by the \dm mass. Scalar, dilaton \dm would cause oscillations in the electron mass or other fundamental constants \cite{Stadnik2015a,Stadnik2015b,Stadnik2016}, resulting in a change of the size of an object \cite{grote2019novel}; axions \cite{Preskill:1982cy} would alter the phase velocity of light \cite{nagano2019axion,nagano2021axion}, affecting the roundtrip time of laser light in LISA; dark photons could couple to baryons or baryon-lepton number in the test masses, leading to a sinusoidal force exerted on them \cite{Pierce:2018xmy}.  While the physics of each model is different, the observable is the same: time-varying positions or accelerations of test masses relative to one another. 

The mass range to which space-based \gwh detectors are sensitive is a function of the frequency range: $\mathcal{O}(10^{-5}-1)$ Hz corresponds to $\mathcal{O}(10^{-19}-10^{-14})$ eV/$c^2$. At such low masses -- compared to those to which ground-based \gwh detectors such as LIGO \cite{2015CQGra..32g4001L}, Virgo \citep{2015CQGra..32b4001A}, and KAGRA \cite{Aso:2013eba} could probe -- the expected signal is monochromatic, since the frequency shift canonically introduced by the motion of the earth through the \dm ``wind'' is very small compared to the frequency resolution of a search. Therefore, we cannot differentiate between a monochromatic and a ``quasi''-monochromatic \uldmh signal at most frequencies to which \lpf and LISA are sensitive\footnote{This depends on the duration of data analyzed, as will be explained in section \ref{sec:dmints}, but is generally true for the expected lifetime of LISA.}.

There is a growing interest in using \gwh detectors to search for \uldm. GEO600 data \cite{Dooley:2015fpa} was recently analyzed using a Logarithmic frequency axis Power Spectral Density (LPSD) method \cite{trobs2006improved,trobs2009improved}, yielding strong constraints on scalar, dilaton dark matter coupling to the beam splitter \cite{Vermeulen:2021epa}. Furthermore, constraints on vector dark matter, i.e. dark photons, were placed using data from the first \cite{Guo:2019ker} and third \cite{LIGOScientific:2021odm} observing runs of advanced LIGO/Virgo/KAGRA that surpassed upper limits from the Eöt-Wash \cite{Schlamminger:2007ht} and MICROSCOPE \cite{Berge:2017ovy} experiments by a few orders of magnitude at frequencies between $\sim100-1000$ Hz ($4\times 10^{-13}-4\times 10^{-12}$ eV/$c^2$). Finally, methods to search for axions and dilatons in different interferometer channels \cite{nagano2019axion,Hall:2022zvi}, for vector bosons with KAGRA \cite{Michimura:2020vxn}, and tensor bosons \cite{Armaleo:2020efr} in LIGO/Virgo/KAGRA and pulsar timing arrays \cite{Armaleo:2020yml} have been developed for such kinds of searches.

What has not yet been done in this field of direct \dmh searches with \gwh detectors is to apply these methods to space-based instruments. While projected sensitivities have been estimated for a variety of \dmh models \cite{arvanitaki2015discovering,Pierce:2018xmy,Morisaki:2020gui}, the development of algorithms tuned towards monochromatic signals in LISA, DECIGO \cite{Kawamura:2020pcg}, and TianQin \cite{Luo:2015ght} has not yet followed. In this work, we take a first step in this direction by performing a search of \lpf data, a precursor to the kind of data that we expect in LISA, for \uldm, which requires that we develop methods to handle specific problems that \lpf faced and LISA will face, such as non-Gaussianities, gaps, and sparse sampling at low frequencies \cite{Ferraioli:2011am,edwards:2020tlp,Baghi:2021tfd} Specifically, we adapt methods developed in the context of quasi-monochromatic, persistent signals emitted by isolated neutron stars \cite{riles2017recent}, planetary- and asteroid-mass primordial black hole binaries \cite{Miller:2020kmv,Miller:2021knj,Guo:2022sdd}, depleting boson clouds around black holes \cite{d2018semicoherent,isi2019directed}, and \dm that couples to ground-based \gwh interferometers \cite{Pierce:2018xmy,Miller:2020vsl}, to look for \dm in space with \lpf. 

The methods presented here are generically good at finding quasi-monochromatic signals in any dataset, regardless of the underlying physics \cite{riles2017recent,sieniawska2019continuous}. They are also (1) robust against noise disturbances such as powerful lines, (2) efficiently deal with gaps in data collection, and (3) are computationally cheap compared to matched filtering, the optimal method to find weak signals buried in noise. In space-based detectors, \gws from a variety of astrophysical sources, such as galactic white dwarf binaries or black hole inspirals \cite{wyithe2003low,Babak:2017tow}, will also be quasi-monochromatic and last for durations longer than the observing time \cite{Miller:2020vsl,Miller:2022wxu}, or at least for greater lengths of time than in ground-based detectors, of $\mathcal{O}$(hours-days).  Currently, proposals for parameter estimation and matched-filtering searches for \gws with LISA struggle with each of the three aforementioned points \cite{Baghi:2019eqo,Dey:2021dem,Blelly:2021oim}; therefore, the work presented here has much farther reaching implications than ``just'' a search for \dm; it represents a comprehensive analysis scheme that can be applied to \emph{any} quasi-monochromatic signal embedded in imperfect, non-Gaussian, gapped LISA data.

This paper is organized as follows: in section \ref{sec:dmints}, we describe the \dmh signal expected at \lpf when it flew; in section \ref{sec:data}, we explain which data segments from the $\sim 1.5$ year run we analyze. Section \ref{sec:meth} focuses on the methods that we use to search for \uldm, one that breaks the data into smaller, Gaussian chunks, and another that match filters the data with a signal model coherently. We present our results in section \ref{sec:results}, which include rejecting strong outliers in the data and upper limits on the strength of the coupling of \dm to baryons. Finally, we draw some conclusions in section \ref{sec:concl} and discuss opportunities for future work. 

\section{Dark matter signal}\label{sec:dmints}

Dark matter could be composed of spin-1 particles, which we denote as dark photons. The relic abundance of dark matter can be explained entirely by dark photons, which could arise from the misalignment mechanism \cite{nelson2011dark,arias2012wispy,graham2016vector}, parametric resonance or the tachyonic instability of a scalar field \cite{agrawal2020relic,Co:2018lka,Bastero-Gil:2018uel,Dror:2018pdh}, or from cosmic string network decays \cite{long2019dark}. Dark photons could couple directly to baryon or baryon-lepton number in the two \lpf test masses, and exert a ``dark'' force on the them, causing quasi-sinusoidal oscillations \cite{Pierce:2018xmy,Miller:2020vsl}. 

Similarly to the ordinary photon, the vector potential for a single dark photon particle can be written as:

\be
\vec{A}(t,\vec{x})=\left(\frac{\hbar \sqrt{2\rhoDM}}{m c^2}\frac{1}{\sqrt{\epsilon_0}}\right)\sin\left(\omega t -\vec{k} \cdot\vec{x}+ \phi\right),
\label{numI}
\ee
where $\omega = 
(m c^2) / \hbar$ is the angular Compton frequency, $\vec{k} = 
(m\vec{v}_{\text{obs}} )/ \hbar $ is the wave vector, $m$ is the mass of the vector field, $\epsilon_0$ is the permittivity of free space and $\phi$ is a random phase.

The Lagrangian $\mathcal{L}$ that characterizes the dark photon coupling to a number current density $J^\mu$ of baryons or baryons minus leptons is:

\begin{equation}
    \mathcal{L} = -\frac{1}{4\mu_0} F^{\mu \nu} F_{\mu \nu} + \frac{1}{2\mu_0} \left(\frac{m c }{\hbar}\right)^2 A^\mu A_\mu - \epsilon e J^\mu A_\mu, \label{lagrangian} \\
\end{equation}
where $F_{\mu\nu}$ indicates the \emph{dark} electromagnetic field tensor, $\mu_0$ is the magnetic permeability in vacuum, $A_\mu$ is the four-vector potential of the dark photon, $e$ is the electric charge, and $\epsilon$ is the strength of the particle/dark photon coupling normalized by the electromagnetic coupling constant.

Dark photons would cause small motions in each of the test masses, and lead to an observable effect because the test masses are separated from each other and hence experience slightly different dark photon dark matter phases. Such a phase difference leads to a change of the arm length over time. What is needed, therefore, is very precise measurements of the positions of the two test masses, something which \lpf provides.

Each test mass experiences an almost identical acceleration:

\begin{equation}
 \vec{a}(t,\vec{x})\simeq\epsilon e \frac{q}{M}\omega \vec{A}\cos(\omega t -\vec{k} \cdot\vec{x}+ \phi), 
\label{accel}
\end{equation}
where $q/M$ is the charge-to-mass ratio of the test masses. 
This effect is in fact a residual one: if the test masses were made of different materials, then the signal induced from dark photons coupling to baryon-lepton number would be enhanced \cite{Michimura:2020vxn}. A simple relation between dark photon parameters and the effective strain $h_D$ experienced by \lpf can be written as \cite{Pierce:2018xmy,Morisaki:2020gui,LIGOScientific:2021odm}:

\bea
\sqrt{\left\langle h_{D}^{2}\right\rangle} &=&C_{\rm LPF}\frac{q}{M}\frac{v_{0}}{2\pi c^{2}} \sqrt{\frac{2\rho_{\mathrm{DM}}}{\epsilon_{0}}}\frac{e\epsilon}{f_{A}}, 
\label{eqn:h000}
\eea
where $f_A=\omega/(2\pi)$ is the frequency of the \dmh particle, and $C_{\rm LPF}=1/3$ is a geometrical factor obtained by averaging the acceleration over all possible dark photon propagation and polarization directions for a single arm, using the appendix in \cite{Pierce:2018xmy} to aid with this calculation. We note $C_{\rm LPF}$ is a factor of $\sqrt{2}$ smaller than the geometrical factor in LIGO $C_{\rm LIGO}=\sqrt{2}/3$, which indicates that an L-shaped interferometer would observe a signal $\sqrt{2}\sim 1.4$ times stronger than a single arm would. 

As we can see from equation \ref{accel}, the test masses will experience a sinusoidal oscillation of their positions over time, at a frequency set by the mass of the \dmh particle. However, if we observe for longer than the \dmh signal coherence time $\tcoh$, then we will resolve stochastic frequency fluctuations $\Delta f$ about the \dmh mass, induced by the motion of the spacecraft relative to the \dmh field. In other words, we wish to contain these fluctuations to one frequency bin $\delta f$, so we require:

\be
 \Delta f=\frac{1}{2}\left(\frac{v_0}{c}\right)^2 f_A < \delta f = \frac{1}{\TFFT} \label{eqn:deltafv}
\ee
which leads to:
\be
\TFFT < 10^7\text{ s} \left(\frac{0.1 \text{ Hz}}{f_A}\right) \sim 
\tcoh \label{eqn:tfft}
\ee
where $\TFFT$ is the fast Fourier transform length, and $v_0\simeq 220$ km/s is the velocity at which dark matter orbits the center of our galaxy, i.e. the virial velocity \cite{smith2007rave}. Here, we observe for a duration much less than $\tcoh$, which implies that the signal will be fully contained within one frequency bin, i.e. it is purely monochromatic for our purposes.

\section{Data}\label{sec:data}

We analyze eight periods of differential acceleration time-series \lpf data, whose details are given in table \ref{tab:segs}, each of which lasts from $\sim 5$ days to $\sim2.5$ weeks \cite{lpfarchive}. One segment, \#6, that began on 14 February 2017, was obtained at a much lower temperature than the others, which resulted in a different noise level \cite{Armano:2018kix}. We therefore treat it as a separate dataset from the others, as will be detailed in Sec. \ref{sec:meth}, and present all constraints based on this dataset.

In table \ref{tab:segs}, we also report the $\TFFT$ such that the data within that segment remain Gaussian at the 5\% significance level, based on the Kolmogorov-Smirnov test \cite{massey1951kolmogorov}. We see here that some segments remain Gaussian for longer compared to others, but in general, the duration of Gaussian data is short compared to the segment duration. Therefore, for the semi-coherent search, we consider $\TFFT=8192$ s for all segments, as explained in the next section.

\begin{table}
\begin{tabular}{|c|c|c|c|c|}
\hline 
number & date & duration (days) & $T_{\rm FFT}$ (s) & Temp (C) \\ 
\hline 
1 & 2016-03-$21$ & 5.08 & 16384 & 22 \\ 
2 & 2016-04-$04$ & 9.33 & 16384 & 22 \\ 
3 & 2016-07-$24$ & 5.29 & 4096 & 22 \\ 
4 & 2016-11-$16$ & 9.82 & 8192 & 22 \\ 
5 & 2016-12-$26$ & 17.86 & 8192 & 22 \\ 
6 & 2017-02-$14$ & 13.33 & 8192 & 11 \\ 
7 & 2017-05-$29$ & 7.04 & 131072 & 22 \\ 
8 & 2017-06-$08$ & 8.4 & 262144 & 22 \\ 
\hline 
\end{tabular}
\caption{\label{tab:segs} Data segments considered in this search obtained from \cite{lpfarchive}. Segment \# 6 is treated as a separate dataset here: we perform coincidences of the other seven datasets with segment \# 6 in order to reduce the number of outliers.}
\end{table}

\section{Method}\label{sec:meth}.

\subsection{Semi-coherent method: projection}

Due to the non-Gaussian nature of the noise when observing for longer than $\sim 8192$ s, we break the dataset into smaller chunks, of duration $\TFFT=8192$ s, analyze these chunks coherently, and then combine their power \emph{incoherently}, that is, without the phase information. To perform this analysis, we leverage existing methods used in continuous-wave data analysis for the search of depleting boson clouds around black holes \cite{d2018semicoherent}, and for \dm that could couple to ground-based \gwh detectors \cite{Miller:2020vsl}.

\subsubsection{Creation of time/frequency peakmaps}

This semi-coherent method \cite{Miller:2020vsl} operates on data in the time/frequency plane, so we take fast Fourier Transforms with $\TFFT=8192$ s, divide the square modulus of the fast Fourier transform by a running-median estimation of the power spectral density to obtain ``equalized power'' (whose mean value in noise should be 1), set a threshold $\theta_{\rm thr}=2.5$ to remove spurious noise peaks, and select local maxima in the power spectra. Each of these choices is motivated in \cite{Astone:2005fj}, and is meant to be robust against the presence of non-stationarities in the noise. This threshold in fact comes from empirical studies in \cite{Astone:2005fj} showing that while the ideal threshold in Gaussian noise is 2, a threshold of 2.5 allows for a higher value of the detection statistic, and a reduction in the number of outliers in the presence of disturbances in real LIGO/Virgo data. The ``local maxima'' criteria is applied because spectral disturbances in real data may not be well localized to one frequency bin (i.e. they have finite coherence times that do not match the $\TFFT$ chosen), or they may turn on and off over the course of the run (or even within one $\TFFT$. Therefore, selecting local maxima helps to minimize contamination of nearby bins from noise lines. The thresholded time/frequency maps are called ``peakmaps'', in which only powerful ``peaks'', i.e. points in the time/frequency plane, remain after the thresholding and local maxima selection. 

This procedure results in the left-hand panel of figure \ref{fig:pm-and-proj}, for segment \#6. We employ the semi-coherent method at frequencies above $\sim 1$ mHz, since below this value, it becomes difficult to estimate the power spectral density using a simple running median, due to the lack of data points at such low frequencies.

\subsubsection{Projection and selection of candidates}

\begin{figure*}[htb]
    \centering
    \includegraphics[width=\columnwidth]{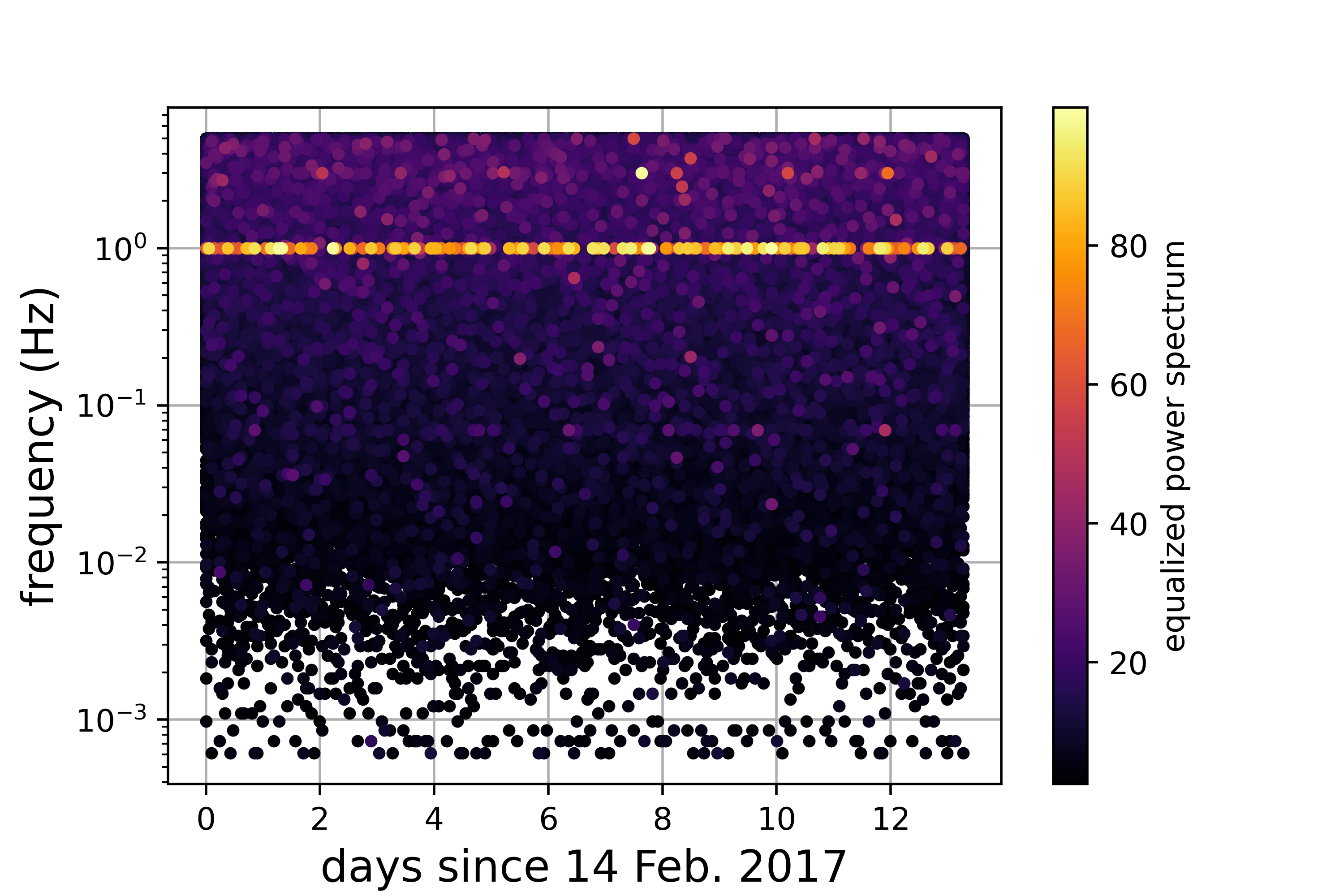}
    \includegraphics[width=\columnwidth]{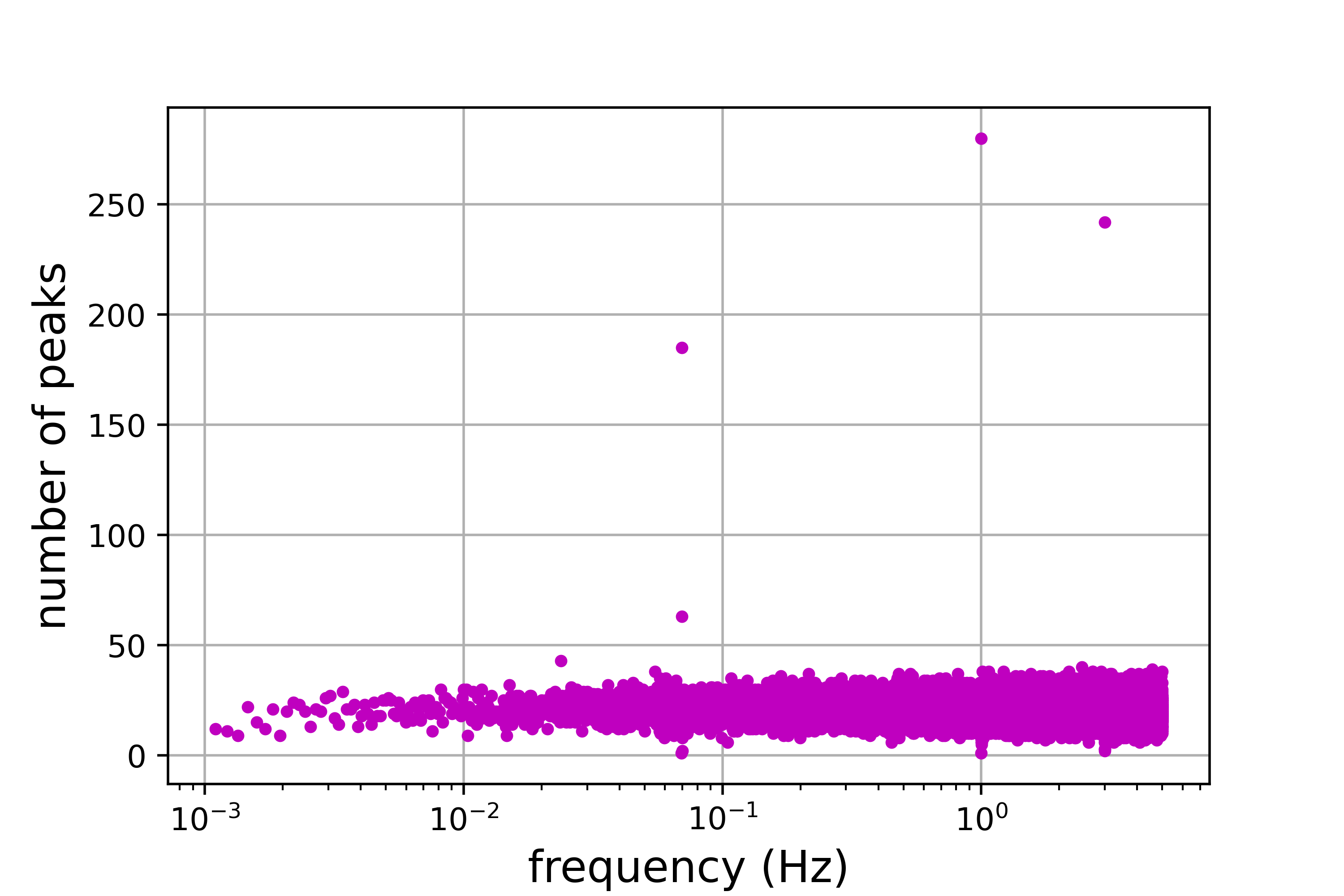}
    \caption{Left: example time/frequency peakmap used in the semi-coherent search. Right: projection of the peakmap onto the frequency axis, corresponding to an integration over time.
    }
    \label{fig:pm-and-proj}
\end{figure*}

After we have created the time/frequency peakmaps, we then attempt to recover some signal power that has been lost due to dividing the data into chunks. To do this, we integrate the peakmap over time, i.e. we project it onto the frequency axis, summing only whether or not a peak appears at a given time and frequency, \emph{not} the equalized power that is given in the color bar of the left-hand panel of figure \ref{fig:pm-and-proj}. This is a choice that is, again, motivated by the presence of noise artifacts: in pure Gaussian noise, summing signal power provides the best chances of detection. However, there are powerful noise lines, e.g. the one at 1 Hz, that we do not want to blind us to possible signals. In this way, we reduce the signal-to-noise ratio of the instrumental lines, while preserving sensitivity towards weak, monochromatic signals. We sum each peakmap from datasets 1-5; 7-8 together in this method, allowing us to recover some signal-to-noise ratio that is lost by breaking the data into chunks of length $\TFFT$.

At this point, we obtain a ``histogram'' of the number of peaks at a given frequency, in the right-hand panel of figure \ref{fig:pm-and-proj}. It is on this projected peakmap that we select possible significant candidates, i.e. particular frequencies, whose number counts are high relative to other frequencies. Since the frequency range spans three orders of magnitude, we select candidates uniformly in the log of frequency. We decide on a certain number of candidates $N_{\rm cand}$ to select such that we would expect a certain number of coincidences $N_{\rm coin}$ between the combined dataset (segments \#1-5; 7-8) and segment \#6 if the data were purely Gaussian in the frequency band of width $B$:

\begin{equation}
N_{\rm cand} \approx \sqrt{N_{\rm coin}\TFFT B}.
\end{equation}
Here, we set $N_{\rm coin}=1$, $B$ = 5 Hz, and $\TFFT=8192$ s, so we select $N_{\rm cand}\approx 200$ candidates uniformly in logarithmic frequency. We select the strongest candidate in each sub-band.

Once we have these candidates, we compute our detection statistic, the critical ratio (CR):

\begin{equation}
CR = \frac{n - \mu}{\sigma},
\end{equation}
where $n$ is the number of peaks at a given frequency, and $\mu$ and $\sigma$ are the average number and standard deviation of peaks across the sub-band, respectively. The CR is a random variable with mean 0 and standard devitation of 1; therefore, we set a threshold $CR_{\rm thr}=5$, corresponding to 5 standard deviations from the mean, that indicates that a particular candidate is ``significant'' and needs to be studied further. 

\subsection{Fully-coherent method: matched-filter}

For the frequencies considered in this analysis, and for the durations analyzed (see table \ref{tab:segs}), equation \ref{eqn:tfft} indicates that the signal will be purely monochromatic; hence, we also perform a fully coherent search of each segment, by simply taking a fast Fourier Transform and computing the so-called matched filter signal-to-noise ratio (SNR) $\rho$:

\begin{eqnarray}
\rho^2&=& 4\text{ Re}\left[ \int_0^{\infty}df\frac{\tilde{h}(f)\tilde{d}(f)}{S_n(f)}\right] \\
&=& 4\text{ Re}\left [ \frac{\tilde{d}(f)^2}{S_n(f)}\right] \label{eqn:snr}
\end{eqnarray}
where the tilde denotes the Fourier Transform of the corresponding quantity, $\tilde{d}(f)$ is the Fourier transform of the time-series $\Delta g$ acceleration data $d(t)$, $\tilde{h}(f)$ is the Fourier Transform of the waveform for the dark-photon signal $h(t)$, and $S_n$ is the power spectral density of the noise. To pass from the first to the second line in the above equation, we note that our desired signal is purely monochromatic. Therefore, the best filter, at each frequency, is simply a monochromatic signal at that frequency, and the number of templates used is simply equal to the number of frequencies analyzed, $N_f\sim3.4\times 10^6$. We impose a threshold $\rho_{\rm thr}=8$ to indicate ``significant'' events, which corresponds to a false alarm probability, in pure Gaussian noise accounting for the trials factor, of 1 per $10^9$.

\subsubsection{PSD estimation at low frequencies}

The sparseness of \lpf data at low frequencies (defined here as less than 1 mHz) is problematic for a running-median estimation of the PSD. Hence, we employ a method developed in \cite{Castelli:2020zro} in order to obtain a robust low-frequency estimation of the PSD. The concept is to average Black-Harris windowed, 50\% interlaced power spectra, obtained with a fixed $\TFFT$ at a given frequency, and then subsequently decrease $\TFFT$ at higher frequencies, resulting in more spectra to average. The frequencies $f_j$ at which the PSD estimation are performed are given by a recursive formula:

\begin{eqnarray}
f_j &=& \frac{2}{(3/5)^{j-1}}f_0 \\
f_0 &=& \frac{M}{N_{\rm max}\Delta T}
\end{eqnarray}
where $M$ is the spacing between frequency bins needed to avoid spectral leakage from the averaging window among neighboring frequencies, $N_{\rm max}$ is the maximum number of samples to fast Fourier transform (for the starting $\TFFT=T_0=2\times 10^5$ s), and $\Delta T=1/10$ s is the sampling time. Here, $N_{\rm max}=\frac{T_0}{\Delta T} =2\times 10^6$ samples, and $M$ is set to 4 bins at $f_0$, and 8 bins for all others, obtained by ensuring a lack of correlation between neighboring bins \cite{Vitale:2014cea}. The factor $3/5$ is obtained by avoiding correlations between two neighboring frequency bins:

\begin{equation}
\frac{M-\alpha}{M+\alpha} = \frac{3}{5},
\end{equation}
where $\alpha=2$ ensures that no correlations in the PSD exist between frequencies spaced by with a spacing $\delta f = \frac{2\alpha}{\TFFT}$.

\begin{figure}
    \centering
    \includegraphics[width=0.49\textwidth]{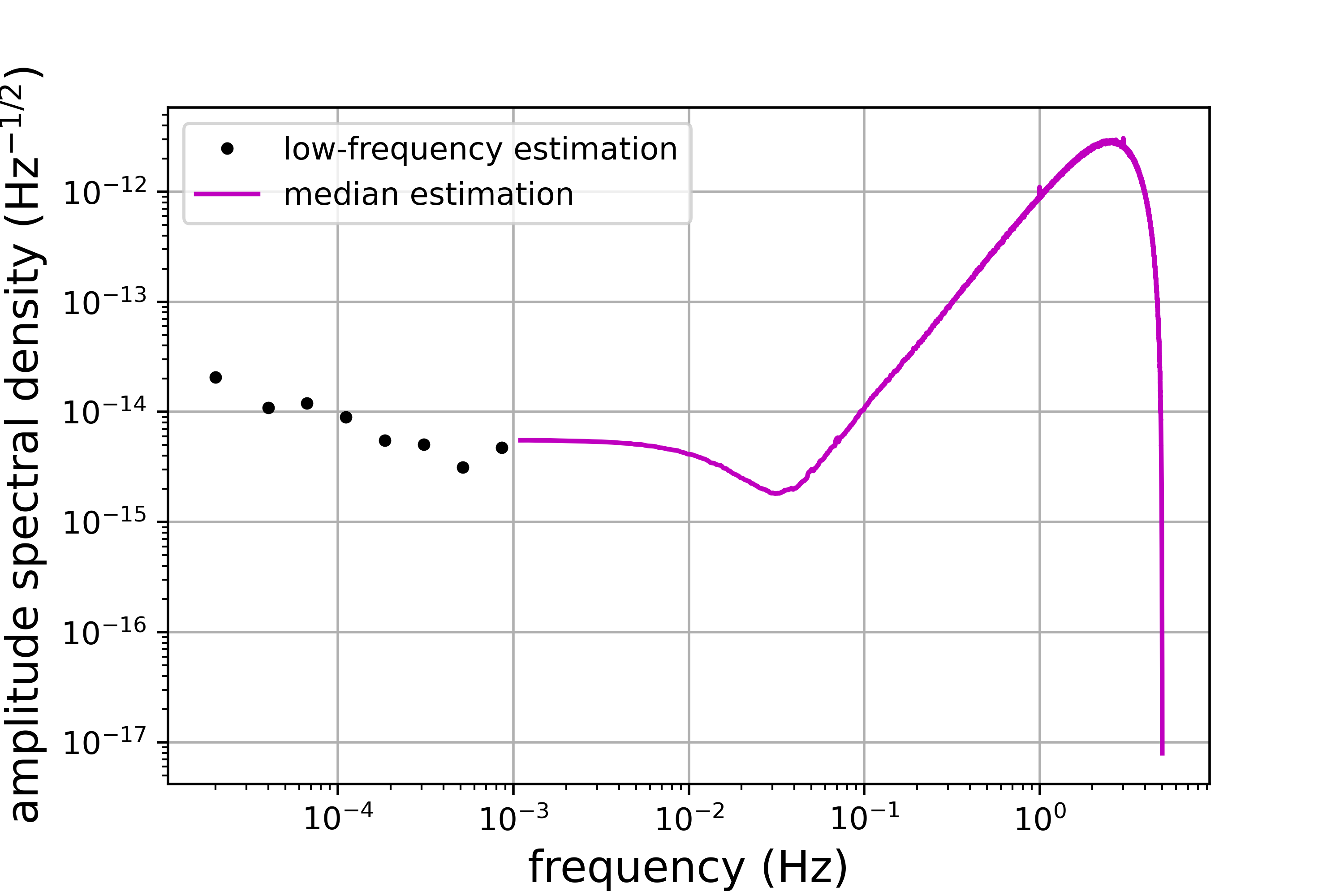}
    \caption{Estimation of the amplitude spectral density at low and high frequencies, using the method described in \cite{Castelli:2020zro} and a running median, respectively.}
    \label{fig:asd-low-high-estimation}
\end{figure}

$\TFFT$ scales with the same relation as:

\begin{equation}
T_{\text{FFT},j}= \left(\frac{3}{5}\right)^{j-1} N_{\rm max} \Delta T
\end{equation}
By windowing, Fourier transforming and averaging the data in chunks of length $T_{\text{FFT},j}$, we obtain the values of the PSD at frequencies between $20\mu$Hz and $\sim 1$mHz, shown in figure \ref{fig:asd-low-high-estimation} as black dots. For the rest of the parameter space, a running median estimation over 20 bins is employed to estimate the PSD, as in many LIGO/Virgo/KAGRA searches.

\subsection{Coincidences}

We have indicated in table \ref{tab:segs} eight segments that have been analyzed in this work. However, we only have data from one detector. Typically, in \gwh data analysis, we look for similar signals in a collection of detectors to rule out the possibility of false alarms. In this case, though, we can look for coincidences between two datasets if their noise distributions are sufficiently different, such that they ``function'' as independent probes of dark matter. Furthermore, the \dmh signal should always be present in the data; therefore, if a candidate is seen in one segment but not in another, this would indicate that it is due to something artificial, not astrophysical.

In the semi-coherent case, we sum the peakmaps from segments $\#1-5; 7-8$ and require that a candidate in one detector appear within 2 frequency bins of a candidate in another. This choice of 2 bins is in fact very generous: the \dmh signal should be \emph{exactly} at the same frequency in each segment. We then impose a threshold on the critical ratio, requiring that a candidate's CR is greater than $CR_{\rm thr}=5$. 

In the fully-coherent searches, we do coincidences between each segment and segment $\#6$, and again require that candidates be within 2 frequency bins of each other. In this case, however, the size of the frequency bins of each segment will be slightly different, since their durations are not equal. We therefore use the larger of the two frequency bins to determine whether a coincidence has occurred. Then, we impose that the SNR must be greater than $\rho_{\rm thr}=8$. As a last stringent check, we will also require that a candidate be present in each data segment for it to be considered as an ``outlier'' worthy of further study.
   
\section{Results}\label{sec:results}

We present here the results of our search for \dm interacting with the test masses in \lpf. 

In the semi-coherent search in the high-frequency regime, the only coincident outliers above $CR_{\rm thr}$ were at $\sim 70$ mHz, 1 Hz and 3 Hz. These frequencies are contaminated with very strong noise lines and can thus be discarded.  At 70 mHz, there is a known noise disturbacne due to the thrusters 
\cite{LISAPathfinder:2019eny}; at 1 Hz and 3 Hz, these are harmonics arising from electrical cross-couplings from the pulse-per-second timing signal present
on the spacecraft \cite{Armano:2022ose}.

In the fully-coherent search, across all frequencies, we only obtained three coincident outliers in at least two of the segments analyzed that were not due to a particular known noise disturbance. They are given in table \ref{tab:cands}, and are barely above the SNR threshold set. We note, however, that these outliers did not appear in each segment that we separately analyzed, indicating that they are due to noise disturbances.

\begin{table}
\begin{tabular}{|c|c|}
\hline 
frequency (Hz) & SNR \\ 
\hline 
0.474766 & 8.57 \\ 
1.213422 & 8.50 \\ 
2.584182 & 8.26 \\ 
\hline 
\end{tabular}
\caption{\label{tab:cands} Remaining candidates from the fully coherent, high-frequency search that appeared in at least two the datasets analyzed. We note that these candidates did not appear in all segments analyzed, as we would expect for a true dark photon signal; thus, we veto them. }
\end{table}

\begin{figure}
    \centering
    \includegraphics[width=0.49\textwidth]{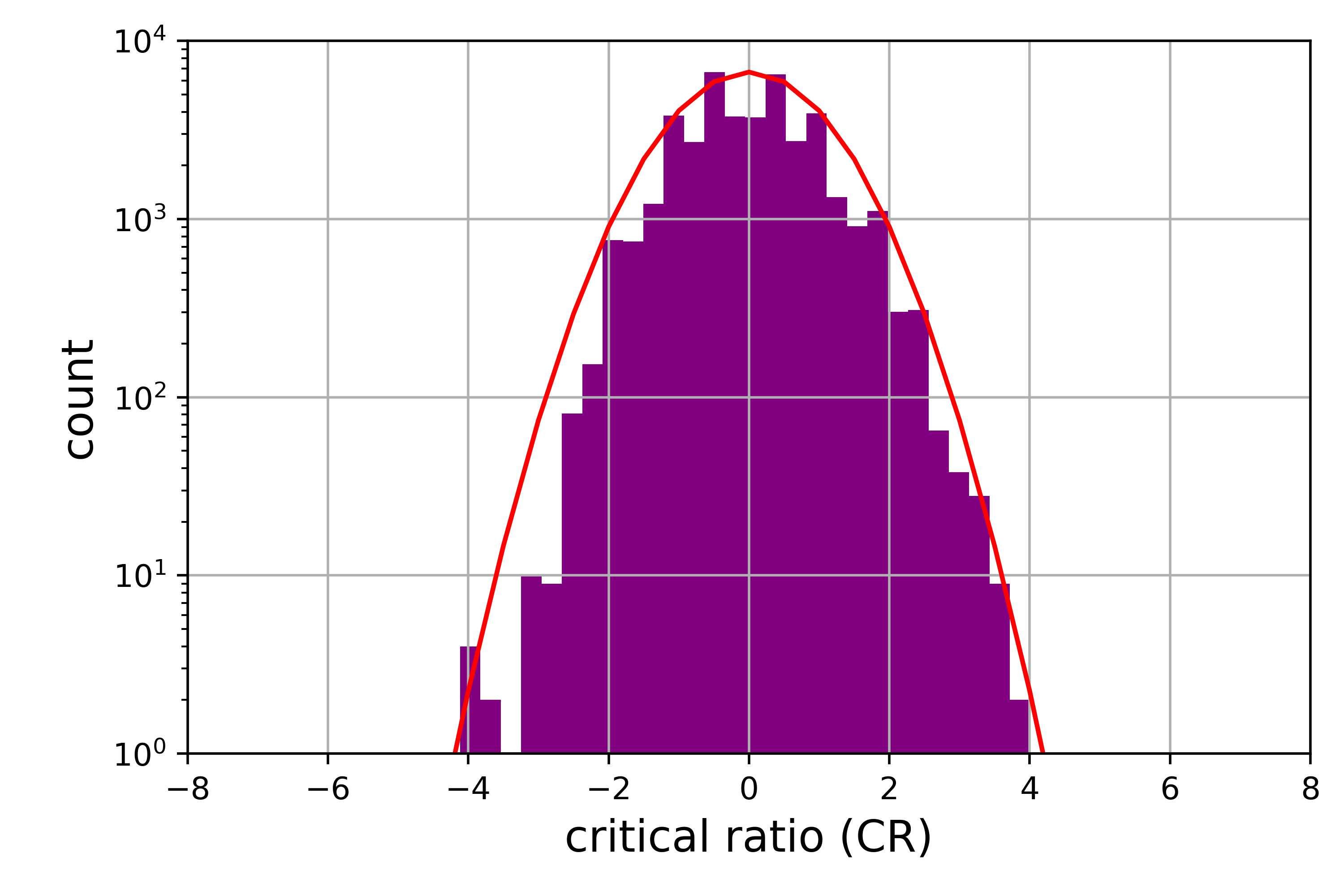}
    \caption{Distribution of the semi-coherent detection statistic $CR$, with a Gaussian curve overlayed for data segment \#6 starting on 14 Feb. 2017.}
    \label{fig:CR-dist}
\end{figure}

\begin{figure}
    \centering
    \includegraphics[width=0.49\textwidth]{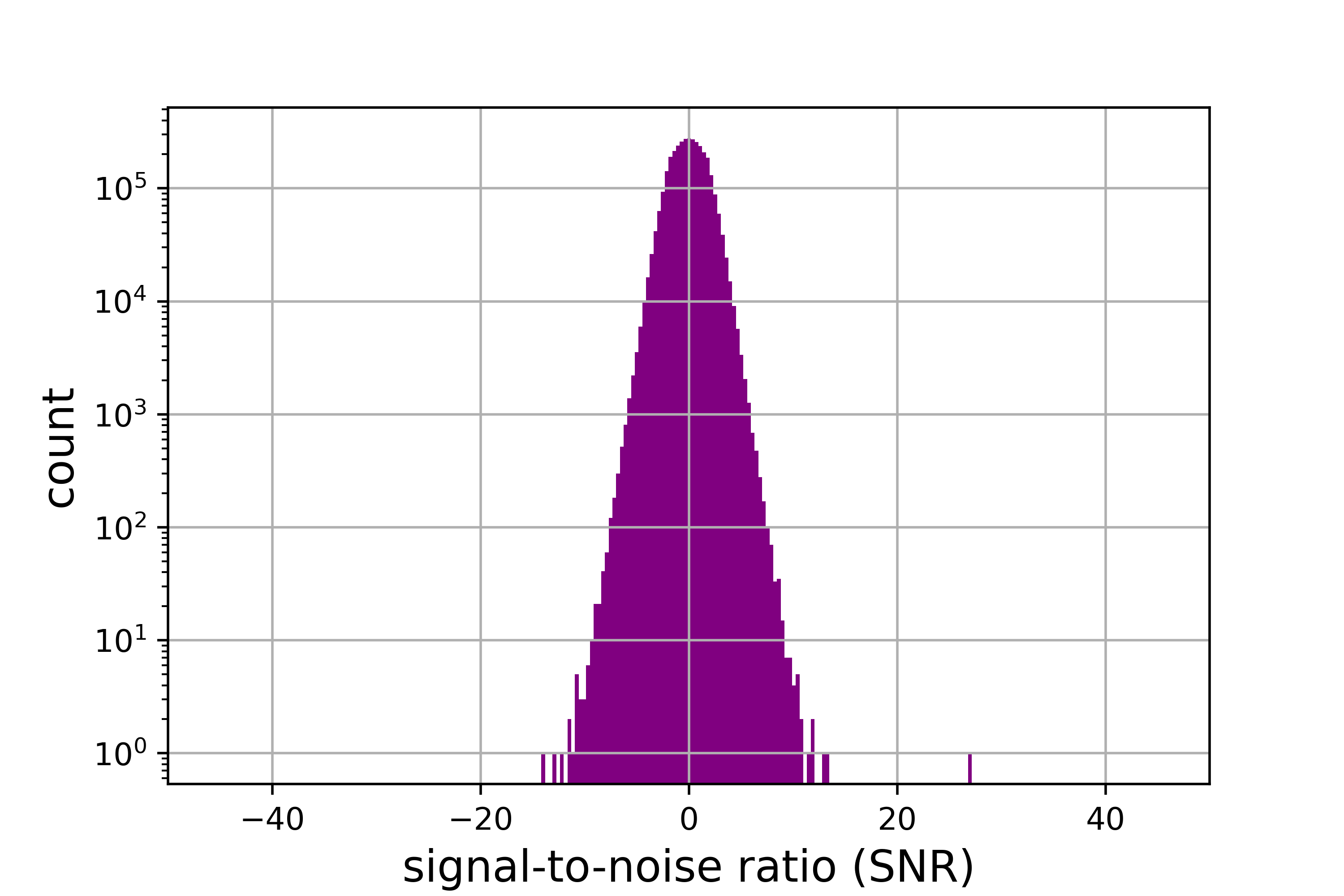}
    \caption{Distribution of the matched-filter detection statistic $\rho$ across the whole frequency range, for data segment \#6 starting on 14 Feb 2017.}
    \label{fig:all-SNR-dist}
\end{figure}

For the critical ratio and the matched-filter SNR, we provide the statistical distributions obtained from the search in figures \ref{fig:CR-dist} and \ref{fig:all-SNR-dist}, respectively. We can see here that these distributions are not exactly Gaussian, due to the presence of non-stationarities in the data. 

\begin{figure}
    \centering
    \includegraphics[width=0.49\textwidth]{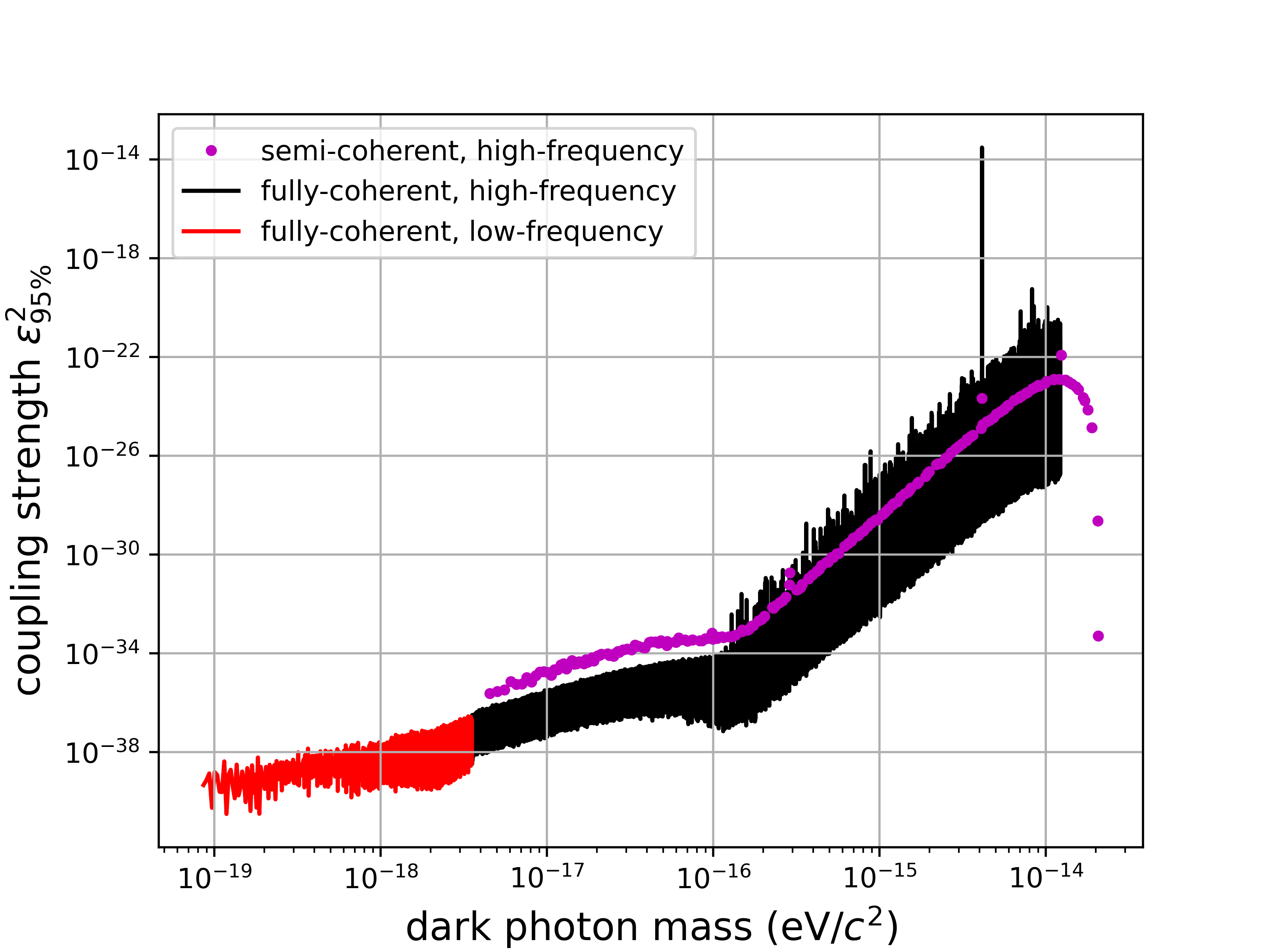}
    \caption{Upper limits in both the fully-coherent and semi-coherent regimes, with the red indicating the regime in which the low-frequency PSD estimation was used, for the period starting 14 Feb. 2017. The mass range shown on the $x$-axis corresponds to the frequency range $[2\times 10^{-5},5]$ Hz.}
    \label{fig:semi-full-all-ul}
\end{figure}

After vetoing all candidates, we then set upper limits on the coupling of \dm to  baryons in the test masses. We provide these limits in figure \ref{fig:semi-full-all-ul} in the low- and high-frequency regime. Specifically, in the high-frequency regime, we calculate these limits with both the results of the projection method and the matched filter for comparison. To obtain these limits, we employ the Feldman-Cousins approach \cite{Feldman:1997qc} to map values of the critical ratio and SNR to ``inferred'' ones to ensure complete coverage at the chosen confidence level (95\% in this case) using table 10 of \cite{Feldman:1997qc}. This procedure inherently assumes that the critical ratio and SNR follow Gaussian distributions, but is robust against non-Gaussianities: it provides conservative results with respect to simulations of dark-photon signals injected in real data (see figure 12 in \cite{Miller:2020vsl}). From these ``inferred'' values of our detection statistics, we compute the constraint on the coupling strength using equation 30 in \cite{Miller:2020vsl}, and equation \ref{eqn:h000} here. Equation 30 in \cite{Miller:2020vsl} arises from theoretically calculating the minimum detectable amplitude of a monochromatic signal that our search could see in Gaussian noise.

We note that the upper limits for the semi-coherent method, which combine the peaks in each segment, are shown for the ``limiting'' segment \#6, since the duration and noise spectral density of this segment impede the sensitivity of the semi-coherent search relative to the other segments. If we were, instead, to use the critical ratios arising from integrating over each of the other segments, the limits would be lower by $\sim 2$, representing the ratio between the total observation time of segments \#1-5; 7-8, and segment \#6.

There is an additional effect that must be accounted for when calculating these upper limits. Due to the fact that $\Tobs<<\tcoh$ for the parameter space considered here, the stochastic nature of the signal amplitude affects the strength of the \dmh coupling that we could observe. In other words, it could be possible that we would observe the 
\dmh field amplitude at a near-zero value, something that does not happen in the regime $\Tobs>>\tcoh$, since we break the data into chunks of length $\TFFT \sim \tcoh$, i.e. we observe for a full coherence time. These effects have been calculated for both scalar, axion \dm \cite{Centers:2019dyn} and vector dark matter (dark photons) \cite{Nakatsuka:2022gaf}. This stochastic effect amounts to an $\mathcal{O}(1)$ loosening of the upper limits in amplitude, and in our case, our limits on $\epsilon^2$ must be increased by a factor of 9. This correction has been applied in figure \ref{fig:semi-full-all-ul}.

\section{Conclusions}\label{sec:concl}

We have performed the first search for dark photon \dm with \lpf data, and have set upper limits on the coupling of dark photons to baryons in the test masses used in this mission. While these limits are not stringent compared to those from torsion balance experiments \cite{Schlamminger:2007ht,Berge:2017ovy}, the work here provides a proof-of-concept pipeline to analyze future LISA data and perform \dmh searches. Interestingly, we had to consider problems such as PSD estimation at very low frequencies, performing coincidences using one instrument, dealing with non-Gaussian artifacts arising from a \gwh detector in space, and breaking the search into semi-coherent and fully-coherent regimes. Even though there were not many visible (by eye) non-Gaussian disturbances, the Gaussianity of this data broke down after $\mathcal{O}(\rm hours-days)$ in each segment analyzed. Our work therefore motivates the development of more advanced PSD estimation algorithms that can handle future data in LISA, the need for semi-coherent, time/frequency-based analyses of the data, and the power that search techniques designed for quasi-monochromatic signals in ground-based \gwh detectors have when applied to a LISA-like mission. 

Furthermore, in LISA, many astrophysical signals will be almost monochromatic -- light binary black hole inspirals, inspiraling galactic white dwarf binaries, etc.--; therefore, our work shows how effective time/frequency analyses, as opposed to pure, computationally expensive matched filters, can be to look for different kinds of signals in future LISA data. While there is of course a sacrifice in sensitivity, evidenced by comparing the purple and magenta curves in the right-hand panel of figure \ref{fig:semi-full-all-ul}, the difference is, at most, an order of magnitude in $\epsilon^2$ at $\sim 1$ mHz, meaning a factor of a few in strain, which is consistent with the comparison of semi-coherent methods to matched filtering given in \cite{Astone:2014esa}.

For \dmh searches, LISA will provide access to masses that ground-based \gwh detectors simply cannot probe, due to seismic activity and Newtonian noise on earth. The sensitivity achievable in LISA is expected to surpass by several orders of magnitude the existing constraints on $\epsilon^2$ from torsion-balance experiments \cite{Pierce:2018xmy}. It is therefore worth investing in data-analysis pipelines in LISA, such as the one employed here, to perform searches to potentially directly detect \dm, and also other astrophysical sources that would emit quasi-monochromatic signals.

Future work will include implementing a more robust estimation of the background of our detection statistics as described in \cite{Tenorio:2021wad}, extending our analysis to the whole \lpf dataset, performing simulations of \dmh particles interacting with the detector, and applying this analysis to other types of \dmh that could be detected, e.g. the dark photon arising from kinetic mixing that couples to the ordinary photon \cite{Chaudhuri:2014dla}. All of these avenues of research will be relevant for when LISA eventually flies.


\section*{Acknowledgments}

LIGO Laboratory which is a major facility fully funded by the National Science Foundation.

Computational resources have been provided by the supercomputing facilities of the Université catholique de Louvain (CISM/UCL) and the Consortium des Équipements de Calcul Intensif en Fédération Wallonie Bruxelles (CÉCI) funded by the Fond de la Recherche Scientifique de Belgique (F.R.S.-FNRS) under convention 2.5020.11 and by the Walloon Region.



We would like to thank Eleonora Castelli for useful discussions on the noise disturbances in Lisa Pathfinder data, and Yue Zhao for a Mathematica code to perform the geometric factor calculation.

This work has made use of the LTPDA MATLAB toolbox.

All plots were made with the Python tools Matplotlib \cite{Hunter:2007ouj}, Numpy \cite{Harris:2020xlr}, and Pandas \cite{mckinney-proc-scipy-2010,reback2020pandas}.

A.L.M. is a beneficiary of a FSR Incoming Post-doctoral Fellowship. We acknowledge support from the ESA Archival Research Visitor Programme, that allowed us to carry out this work.

We would like to thank all of the essential workers who put their health at risk during the COVID-19 pandemic, without whom we would not have been able to complete this work.

\bibliographystyle{apsrev4-1}
\bibliography{lf_analysis}

\end{document}